\documentclass{iaus}
\usepackage{graphicx}

\title[Spatially varying mass function of MACHOs]
{Spatially varying mass function of MACHOs in the galactic halo
and interpretation of microlensing results }

\author[Sohrab Rahvar]%
{Sohrab Rahvar
}

\affiliation{Department of Physics, Sharif University of
Technology,
P.O.Box 11365--9161, Tehran, Iran, email:rahvar@sharif.edu\\[\affilskip]
Institute for Studies in Theoretical Physics and Mathematics,
P.O.Box 19395--5531, Tehran, Iran}

\pubyear{2004}
\volume{225}
\pagerange{1--8}
\date{?? and in revised form ??}
\jname{Impact of Gravitational Lensing on Cosmology}
\editors{Mellier, Y. \& Meylan,G. eds.}
\begin{document}

\maketitle

\begin{abstract}
The gravitational microlensing experiments in the direction of
Large Magellanic Cloud (LMC) predict a large amount of white
dwarfs ($\sim 20\%$) filling the galactic halo. However, the
predicted white dwarfs have not been observed at the galactic
halo. To interpret the microlensing results and resolving the
mentioned problems, we use the hypothesis of spatially varying
mass function of MACHOs, proposed by \cite{ker98} (hereafter KE).
However the KE model is not compatible with the duration
distribution of events \cite{rah04a}. Here we use more realistic
power-law model of MF, $dn/dm\propto m^{\alpha}$ for the MACHOs of
halo. The index of MF in this model changes from $-2.7$ for stars
with $m>1 M_{\bigodot}$ at the central part of galactic halo to
the substellar regime with an upper limit of $-1$ at the edge of
halo. We show that in contrast to the abundant brown dwarfs of
galactic halo, heavy MACHOs can be responsible for the
microlensing events in the direction of LMC.
\end{abstract}

\firstsection 
\section{Introduction}
The rotation curves of spiral galaxies and the Milky Way show that
these type of galaxies have dark halo component. The most trivial
candidate for the dark halo structure is the baryonic matter that
can be in the form of gaseous or MAssive Compact Halo Objects
(MACHOs). Since MACHOs are expected to be too light to be luminous
and difficult detectable, \cite{pac86} proposed an indirect method
so-called gravitational microlensing to observe them indirectly.
Following his suggestion several experiments such as EROS and
MACHO started monitoring millions of LMC stars for one decade and
observed less than 20 events in this direction
(\cite{alc00,las00}). Due to the degeneracy nature of
gravitational microlensing problem, it is impossible to obtain the
mass, distance and transverse velocity of the lenses by measuring
the duration of events. The only way is using statistical studies,
comparing the microlensing events with the models. The outcome of
this study is the mean mass of MACHOs and their mass contribution
in the galactic halo. For the standard Galactic model with the
Dirac-Delta mass function (MF), halo is comprised by $20\%$ of
MACHOs with the mass of $0.5 M_{\bigodot}$. Comparing the luminous
mass of Milky Way with that of halo, we can conclude that the
white dwarfs of halo should have twice mass of the ordinary stars
of the disk and bulge. \\
KE proposed using Dirac Delta spatially varying MF as a solution
to this problem but it has been shown that the duration
distribution in this model is not compatible with the LMC
microlensing candidates. Rahvar (2004a) used a likelihood analysis
to find the best parameters for the model. Here we use more
realistic MF for the MACHOs of halo as the power law function
where the mass scale and the index of MF changes monotonically
from the center to the edge of galaxy. We show that duration
distribution of events in this model is compatible with the LMC
microlensing candidates.
The effective mean mass of lenses, as the mean mass of observed
lenses in this model is larger than the mean mass of overall
lenses of the galactic halo. This means that the galactic halo
with the dominant brown dwarfs can
produce microlensing events compatible with the observed data. \\
The organization of paper is as follows: In Section 2 we introduce
the hypothesis of spatially varying MF of KE and power-law models.
In Section 3 we compare the LMC microlensing events of MACHO
experiment with the expected distribution from the model and
section 4 contains the conclusion of this work.
\section{Spatially Varying MF}
\begin{figure}
\begin{center}
\includegraphics[height=2.5in,width=2.5in,angle=0]{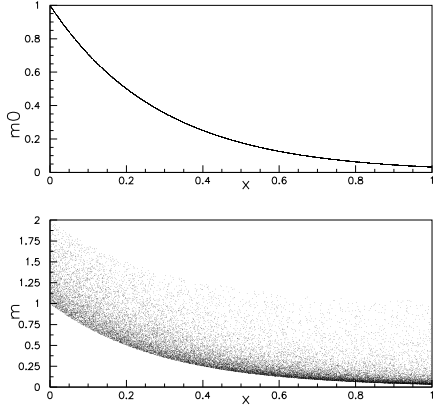}
\caption{The mass distribution of MACHOs as a function of
$x=D_{ol}/D_{os}$, where $D_{ol}$ and $D_{os}$ are the
observer-lens and observer-source distances. The upper panel and
lower panels show the distributions of $m_0$ and mass of MACHOs as
function of distance.} \label{fig:mfdist}
\end{center}
\end{figure}

The tradition in the microlensing analysis is using Dirac-Delta MF
for the MACHOs of galactic halo. This function has been used for
the sake of simplicity to apply it to an arbitrary MF of halo.
However this approach hides the information about the contribution
of the MACHOs with various masses in the gravitational
microlensing candidates.\\
Star and baryonic cluster formation theories (e.g., \cite{ash90};
\cite{car94}; \cite{de95}) predict the variation of MF in
interstellar medium depending on the density of clouds forming the
stars. \cite{shad04} and \cite{elm04} also showed that instead of
universal initial mass function we can have a composite with
various contributions from the brown dwarfs, solar masses and high
mass stars, depending on the density of environment. Since the
halo density changes as $\sim 1/r^2$, we can conclude that the MF
of MACHOs may change by distance from the center of galaxy. The
inner halo comprises partly visible stars, in association with the
globular cluster population, while the
outer halo comprises mostly low-mass stars and brown dwarfs.\\
The hypothesis of using spatially varying MF has been proposed by
KE. They used a spatially varying MF as
\begin{equation}
MF(r)=\delta[M - M(r)],
\end{equation}
where the mass scale in this model decreases monotonically as a
function of $r$ as
\begin{equation}
M_U(\frac{M_L}{M_U})^{r/R_{halo}},
\end{equation}
$M_U$ and $M_L$ are the upper and lower limits for the mass of
MACHOs and $R_{halo}$ is the size of galactic halo. In the CMD
plus baryonic halo models, $R_{halo}$ the size of halo which
contains MACHOs can be extended more than the halo size. \\
Here we use more realistic spatially varying MF than the KE model.
The MF of stars is taken as the power low function $dn/dm\propto
m^{\alpha}$, where the index of function depends on the mass scale
of stars. The upper limit of index \cite{chab02} for the
substellar regime can be $-1$ and the lower limit \cite{sca86} of
$-2.7$ for the stars massive than the solar mass. \\
Here we use the following anzats for the spatially varying MF:
\begin{equation}
\frac{dn}{dm} = [\frac{m}{m_0(r)}]^{\alpha(r)}, \label{pwlmf}
\end{equation}
where $m(r)>m_0(r)$ and the index of power-law changes as
$\alpha(r) = 1.7 \frac{r}{R_{halo}} - 2.7$. The typical mass of
MACHOs, $m_0(r)$ changes as $M_U(\frac{M_L}{M_U})^{r/R_{halo}}$.
Here we take the $M_U = 1 M_{\bigodot}$, $M_L = 0.001
M_{\bigodot}$ and $R_{halo} = 100 kpc$. Figure \ref{fig:mfdist}
shows the distribution of mass of MACHOs as a function of distance
from the center of galaxy.
\section{Microlensing Analysis in the Spatially Varying MF}
\begin{figure}
\begin{center}
\includegraphics[height=2in,width=3in,angle=0]{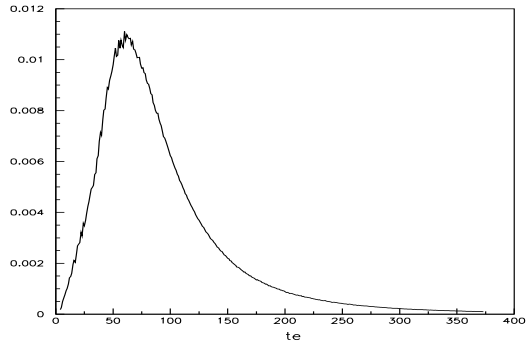}
\caption{The duration distribution of LMC microlensing events
$(t_e)$ , considering the observational efficiency of MACHO
experiment in Standard halo model and spatially varying power-law
MF.} \label{fig:tdist}
\end{center}
\end{figure}

Here in this section our aim is to compare the microlensing data
with the theoretical model. First we calculate the theoretical and
experimental optical depths (\cite{alc00}) and compare them with
each others.
Using the experimental efficiency of observation, the observed
optical depth obtain as follows:
\begin{equation}
\tau_{obs} = \frac{\pi}{4E}\Sigma t_i,
\end{equation}
where $t_i$ is the duration of ith event and $E$ is the exposure
time of observation which for the 5.7 yrs observation of LMC by
MACHO experiment, $E$ is $ 6.12\times 10^7$. The optical depth
results from 13 LMC microlensing candidates is about $4.43\times
10^{-8}$. One the other hand, using the power-law MF for the
MACHOs of galactic halo in the standard halo model, we can produce
the rate of microlensing events $\frac{d\Gamma}{dt}$. The
theoretical optical depth is calculated as
\begin{equation}
\tau_{expected} =
\frac{\pi}{4}\int{\frac{d\Gamma}{dt}\epsilon(t)tdt},
\end{equation}
where we consider hundred percent of halo is made by MACHOs and
$\epsilon(t)$ is the observational efficiency. The ratio of the
observed to the theoretical optical depth results the MACHOs
contribution in the galactic halo, $f
=\frac{\tau_{obs}}{\tau_{expected}}$. Here in this model $23$
percent of halo is made by MACHOs. Figure \ref{fig:tdist} shows
the distribution of duration of events in the standard halo model,
considering the observational efficiency of MACHO experiment.\\
We do a Monte-Carlo simulation to show quantitatively how the
contribution of massive lenses in the microlensing events is more
than the light ones. According to the distribution of matter along
the line of sight the lenses are selected by the probability
function as follows:
\begin{equation}
\frac{dP}{dx}\propto \rho(x)\sqrt{x(1-x)},
\end{equation}
where $\rho$ is the matter density and $x$ is the lens-source to
the observer-source distances. After selecting the position of
lenses, we use the equation (\ref{pwlmf}) to obtain their masses.
The duration of events are generated through $te =
\frac{2R_E}{v_t}$ and compared with the observational efficiency
to be selected or rejected. $R_E$ is the Einstein radius of a lens
and $v_t$ is its transverse velocity with respect to our line of
sight. The mass of selected events are used to obtain the mean
mass of lenses. We called the mean mass of these lenses as the
active mean mass.  In contrast to the active mean mass we define
passive mean mass of lenses as the mean mass of overall lenses in
the halo. The active and passive mean masses of lenses in this
model are obtained as the $0.40$ and $0.05$ solar masses. This
means that brown dwarfs are the abundant MACHOs of halo while
white dwarfs are responsible to the microlensing events.

\section{Comparison with the observed data}
\begin{figure}
\begin{center}
\includegraphics[height=3in,width=3in,angle=0]{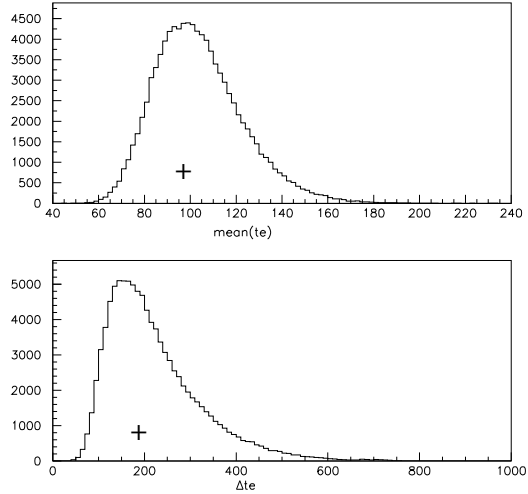}
\caption{Upper and lower panels show the theoretical distributions
of the mean and the width of microlensing events in standard halo
model with spatially varying power-law MF. The cross signs are the
mean and the width of duration of observed LMC candidates of MACHO
experiment.} \label{fig:mwdist}
\end{center}
\end{figure}

In this section we compare the expected duration distribution from
the model with duration distribution of the observed LMC
candidates. Two statistical parameters as the mean and the width
of the duration distribution of events (\cite{gre02};
\cite{rah04b}) are used for our comparison. The width of the
duration distribution for the $N_{obs}$-th observed candidate is
defined by
$$ \Delta t_E = Max_{[j = 1,N_{obs}]}(t_j) - Min_{[j=1,N_{obs}]}(t_j).$$
The mean and the width of duration distribution of MACHO
candidates are $97$ and $188.2$ days. Figure \ref{fig:mwdist}
compares the observed mean and width of duration of events with
what we expect form the theoretical model. We perform a
Monte-Carlo simulation to generate the mean and the width of
distribution of events. Thirteen events is picked up from the
duration distribution of events (Figure \ref{fig:tdist}) and each
time the mean and the width of events are calculated in each set
to produce their distributions. Comparing these distributions with
the observation show that the standard halo model with the
power-law spatially varying MF is in good agreement with the LMC
data.

\section{Conclusion }\label{ap:boundcon}
In this work we use an anzats for the spatially varying power-law
MF for the MACHOs of galactic halo. The advantage of using this MF
is that (i) it is more realistic than Dirac-Delta MF, (ii) the
expected duration distribution of events is in good agreement with
the LMC microlensing data and (iii) the active mean mass of lenses
as the mean mass of lenses contributing in the microlensing events
obtained about $0.40 M_{\bigodot}$ compared to the passive mean
mass of lenses as the mean mass of overall lenses of the galactic
halo with $0.05 M_{\bigodot}$. We showed that in a halo with
abundant brown dwarfs, massive MACHOs with the white dwarf sizes
can be responsible for the microlensing events that we are
observing in the experiment.
\begin{acknowledgments}
I would like to say my thanks to the organizing committee of
IAU225 and physics department of Sharif university, supporting me
at this symposium.
\end{acknowledgments}


\end{document}